\documentclass[journal]{IEEEtran}

\ifCLASSINFOpdf
  \usepackage[pdftex]{graphicx}
\else
\fi

\usepackage{amsmath}
\usepackage{amssymb}
\usepackage{url}
\usepackage{comment}

\usepackage[breaklinks=true,bookmarks=false]{hyperref}

\usepackage{microtype}      % microtypography
\usepackage{xcolor}         % colors
\usepackage{multirow}
\usepackage{graphicx}
\usepackage{amsmath}
\usepackage{caption}
\usepackage{subcaption}
\usepackage{booktabs}
\usepackage{xcolor}         % colors

\begin{document}

\title{A low-complexity method for efficient depth-guided image deblurring}

\author{Ziyao~Yi, Diego~Valsesia, Tiziano~Bianchi, and~Enrico~Magli% <-this % stops a space
\thanks{The authors are with Politecnico di Torino -- Department of Electronics and Telecommunications, Italy. email: \{name.surname\}@polito.it. This study was carried out within the FAIR - Future Artificial Intelligence Research and received funding from the European Union Next-GenerationEU (PIANO NAZIONALE DI RIPRESA E RESILIENZA (PNRR) – MISSIONE 4 COMPONENTE 2, INVESTIMENTO 1.3 – D.D. 1555 11/10/2022, PE00000013, CIG B421A95680, CUP E13C22001800001). This manuscript reflects only the authors’ views and opinions, neither the European Union nor the European Commission can be considered responsible for them.}}% <-this % stops a space

\maketitle

\begin{abstract}
Image deblurring is a challenging problem in imaging due to its highly ill-posed nature. Deep learning models have shown great success in tackling this problem but the quest for the best image quality has brought their computational complexity up, making them impractical on anything but powerful servers. Meanwhile, recent works have shown that mobile Lidars can provide complementary information in the form of depth maps that enhance deblurring quality. In this paper, we introduce a novel low-complexity neural network for depth-guided image deblurring. We show that the use of the wavelet transform to  separate structural details and reduce spatial redundancy as well as efficient feature conditioning on the depth information are essential ingredients in developing a low-complexity model. Experimental results show competitive image quality against recent state-of-the-art models while reducing complexity by up to two orders of magnitude.
\end{abstract}

\section{Introduction}
\label{sec:introduction}
In recent years, mobile devices such as smartphones and tablets have become the prime tools for capturing pictures, with the ability to capture and display high-quality images already becoming a measurement of the equipment’s quality. Learning-based image processing algorithms such as deblurring \cite{kupyn2018deblurgan,liu2024deblurdinat}, denoising \cite{chen2022simple,milanfar2024denoising}, and image super-resolution \cite{gao2023implicit,lu2022transformer}, are being increasingly used to meet the demand of customers for image quality. However, current literature on the topic is mostly focused on achieving the highest image quality, with little attention to computational complexity, particularly for deployment on resource-constrained edge devices.  While some efficient approaches have been proposed to address these challenges \cite{li2023ntire,kong2023efficient,mansour2023zero}, their performance is often limited by the inability of lightweight models to extract enough meaningful information from a single data modality.

As mobile technology continues to advance, newer devices are being integrated with multimodal imaging platforms, which combine multiple kinds of imaging devices. In particular, Lidar sensors are widely equipped in mobile devices such as Apple's iPhone \cite{applelidar}. These sensors provide depth information by sending a grid of light pulses and measure the return time to estimate the distance at multiple points. Since such depth data is sourced from an active instrument, it can offer complementary spatial and structural information that is not affected by motion blur and can be valuable for the image deblurring task, as recently shown in \cite{yi2024deep}. While \cite{yi2024deep} showed the effectiveness of mobile Lidar depth maps for regularization of image deblurring, and introduced some key ingredients towards efficient models, namely efficient adapters and continual learning techniques, the overall computational complexity of the presented models remained high for edge devices.

In this paper we design a low-complexity model, called EDIBNet (Efficient Depth-guide Image Debluring network), for depth-guided deblurring that can be run efficiently on edge devices thanks to a reduction of up to two orders of magnitude in FLOPs, runtime and memory requirements, while maintaining an image quality close to that of state-of-the-art models. Our solution relies on the discrete wavelet transform (DWT) to provide a scale-space decomposition into sub-bands composed of low- (LL) and high-frequency (LH, HL, HH) components. This representation allows us to significantly reduce both runtime and computational cost since deblurring operations can be limited to the first-level decimated LL sub-band, skipping high-frequency details in the other sub-bands that have minimal impact on the dublurring performance. This is integrated into a lightweight neural network with an encoder-decoder architecture. Additionally, real-world depth information, acquired from mobile LiDAR, is fed into the network through efficient lightweight adapter modules that effectively fuse spatial features from both modalities. This design allows the network to benefit from geometric priors while remaining computationally practical for edge deployment.

To summarize, our contributions are as follows:
\begin{itemize}
    \item We design a compact encoder-decoder architecture tailored for edge deployment, which seamlessly integrates real-world depth information from mobile LiDAR via a lightweight adapter. This fusion enhances structure-aware restoration while maintaining low computational overhead.

    \item  We employ the discrete wavelet transform to decompose images into sub-bands, enabling efficient representation and computation while preserving both global structures and local details.
    
    \item Our model processes a high-resolution frame in only 0.2 seconds on an NVIDIA Jetson Orin Nano edge device, showing a reduction of two orders of magnitude with respect to state-of-the-art models while providing competitive image quality.
\end{itemize}

\section{Background and Related Work}
\subsection{Image deblurring}

Deblurring is a classic ill-posed inverse problem, whose aim is to reconstruct a sharp image from blurred observations. Mathematically, the observed image $\mathbf{y}$ is expressed as the convolution of the original image $\mathbf{x}$ with a blur kernel $\mathbf{k}$:
\begin{align}
    \mathbf{y} = \mathbf{k}  \circledast  \mathbf{x}.
\end{align}
where $\circledast$ denotes convolution.

With the rise of deep learning, learning-based approaches have achieved great success thanks to their ability to learn complex priors directly from training data. A common paradigm is supervised learning, which relies on paired blurred and sharp images during training to optimize deep models.\cite{Zhang_2018_CVPR,Kupyn_2018_CVPR,9008540,dong2023multi,Kim2022MSSNet,chen2021pre,wang2022uformer,zamir2022restormer,chen2022simple}. For example, Zhang et al. \cite{Zhang_2018_CVPR} proposed a network consisting of three CNNs and an RNN for deconvolution. DeblurGAN \cite{Kupyn_2018_CVPR} and DeblurGAN-v2 \cite{9008540} introduced adversarial training for image deblurring. Recent methods such as MRLPFNet \cite{dong2023multi} and MSSNet \cite{Kim2022MSSNet} adopt multi-scale architectures to jointly capture fine details and global structures.

With the success of Transformers in natural language processing \cite{vaswani2017attention} and computer vision \cite{dosovitskiy2020image}, they have also been increasingly applied to deblurring. IPT \cite{chen2021pre} was among the first to use standard Transformer blocks trained on large-scale datasets. Uformer \cite{wang2022uformer} introduced a U-shaped Transformer architecture for efficient image restoration, while Stripformer \cite{Tsai2022Stripformer} used intra- and inter-strip tokens to enhance feature representation. Restormer \cite{zamir2022restormer} demonstrated that long-range dependencies can be captured effectively with improved computational efficiency. Meanwhile, NAFNet \cite{chen2022simple} showed that simplified network designs with gating mechanisms can achieve competitive performance without relying on costly self-attention modules.

In addition to focusing only on blurry images, some works have studied the use of information from other domains such as segmentation maps \cite{hyun2013dynamic,10445844}, optical flow \cite{hyun2014segmentation}, and event-camera data \cite{yang2023deformable} to better regularize the reconstruction process and improve image quality.
A small number of works have also explored using depth information to improve deblurring performance. \cite{6215220} proposed a hierarchical depth estimation based on region trees to progressively generate credible blur kernel estimates. Li et al. \cite{9043904} first extract the depth map and adopt a depth refinement network to restore the edges and structures in the depth map. Apart from these, Yi \cite{yi2024deep} demonstrates that utilizing true depth information generated from mobile Lidar can also significantly boost the effectiveness of deblurring algorithms.

\subsection{Efficient deblurring}
A few works have recently studied more efficient deblurring models. Hu et al \cite{Hu_2021_ICCV} proposed PyNAS, which adopts gradient-based search strategies and innovatively searches the patch and scale hierarchy, achieving  58 fps for 720p images on Titan XP GPU. Ruan \cite{ruan2023revisiting} introduced a unified lightweight CNN network that features a large effective receptive field and demonstrates comparable or even better performance than Transformers while bearing less computational costs scheme not limited to cell searching. Zhang \cite{zhang2020rethinking} utilized a modified U-Net as the backbone and proposed a new self-supervised method that tripled the inference speed. Purohit \cite{purohit2019motion} presented a new architecture featuring spatially adaptive residual learning modules that implicitly detect spatially varying shifts causing non-uniform blur in the input image and learn to modulate the filters accordingly.

To eliminate the computational complexity and improve runtime, in addition to simplifying the network architecture, the other common idea is to reduce the resolution of the input. Reducing the resolution by naively downscaling the input size would cause an irreversible loss of essential information, making it harder to recover the high-quality image. An alternative approach is to divide the image into patches and send them to decoders based on image complexity, which is often predicted by a simple classifier. This kind of methods are already been applied in super-resolution \cite{kong2021classsr,wang2024camixersr} and deblurring \cite{mao2024adarevd}. While these methods offer a good trade-off between performance and efficiency, they typically require training multiple decoders of varying capacity and involve complex patchification and scheduling strategies. This increases the training difficulty and model size, making them challenging to deploy in practice.

\subsection{Wavelet-based image restoration}
The wavelet transform is widely used in image processing because it can efficiently capture both spatial and frequency information in multi-resolution scales. A few works attempted to integrate it in deep learning models. Wavelet-srnet \cite{huang2017wavelet} predicts the low-resolution corresponding series of high-resolution wavelet coefficients before reconstructing HR images. WeConvene \cite{fu2024weconvenelearnedimagecompression} introduces a video compression framework that operates in the wavelet domain by integrating the discrete wavelet transform into convolutional layers and performing quantization and entropy encoding in that domain. In terms of deblurring, Gao et al. \cite{gao2024efficient} propose a multi-scale network with a learnable
DWT. Rao et al. \cite{rao2024rethinking} integrate a diffusion model into the Wavelet-Aware Dynamic Transformer (WADT) to address video deblurring.

\section{method}

\begin{figure*}[t]
  \centering    
  \includegraphics[width=0.9\linewidth]{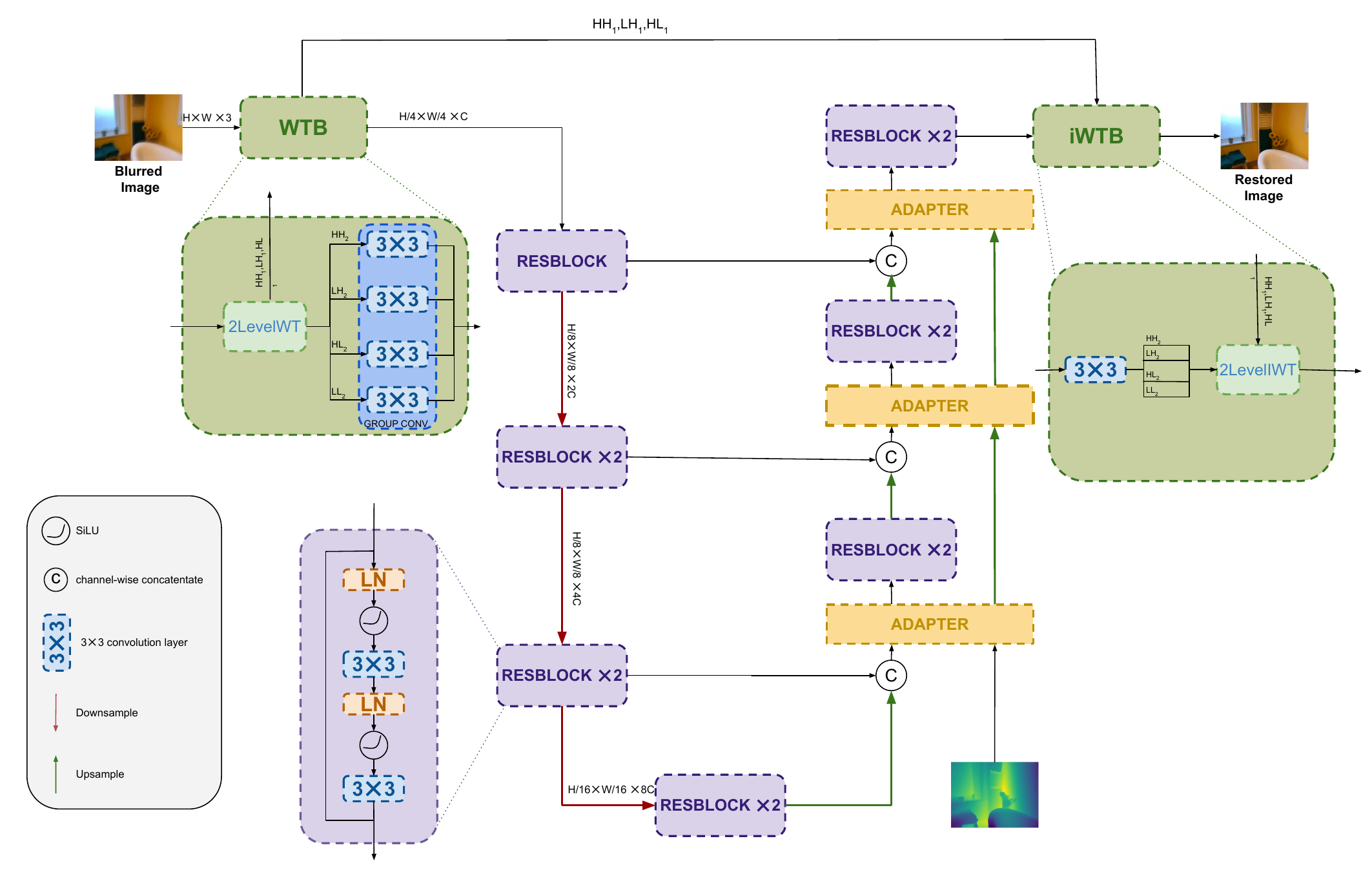}
    \caption{Proposed EDIBNet architecture. An efficient encoder-decoder neural network operates in the low-frequency wavelet sub-bands. Efficient adapters are added on each level of the decoder part to modulate image features with depth features.}
    \label{fig:efficientNetwork}
\end{figure*}

\begin{figure}[t]
  \centering    
  \includegraphics[width=\linewidth]{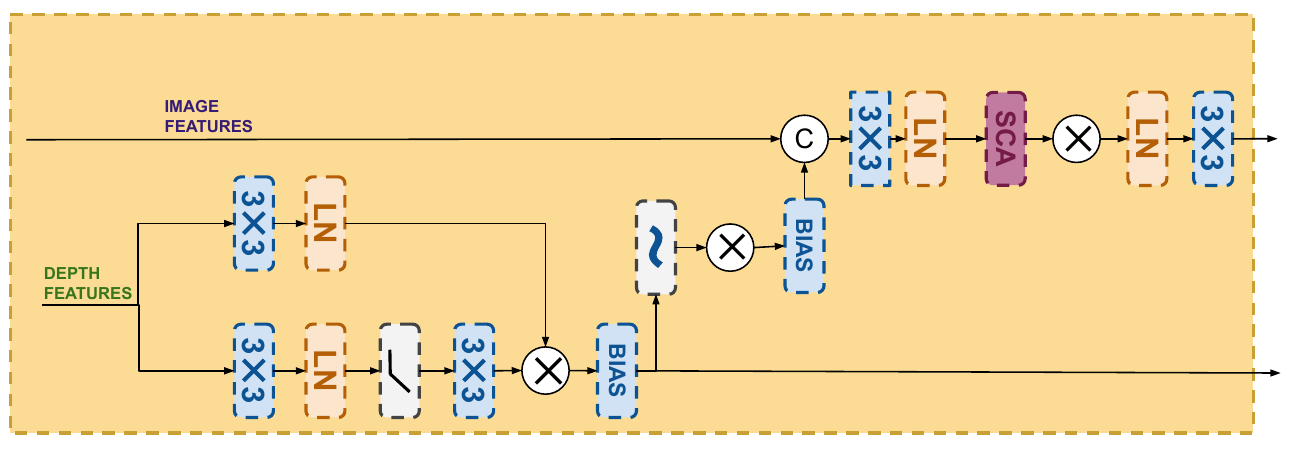}
    \caption{Architecture of the proposed efficient Adapter. The module takes as input both image features and depth features. Each input is first normalized and passed through lightweight bias adjustment layers. The features are then concatenated and processed through a chunking and spatial conditioning mechanism, which generates prompt signals that adaptively modulate the image features based on geometric context. This design enables efficient and structure-aware fusion of depth cues into the visual pipeline with minimal computational overhead.}
    \label{fig:adapter}
\end{figure}

In this section we present the proposed EDIBNet architecture for efficient depth-guided image deblurring. The overall architecture is shown in Fig. \ref{fig:efficientNetwork}. At a high level, the blurred image is first transformed to the wavelet domain and encoded by the Wavelet Transform Block. An encoder-decoder backbone processes the decimated low-frequency sub-bands treated as channels to ensure lightweight operations. Depth information is processed and integrated via efficient adapters in the decoder part of the network. Finally, the image is transformed back to the RGB domain by means of the inverse wavelet transform. Notice that high frequency components are directly skipped to the inverse transform, as they only marginally contribute to improved image quality and their skipping allows to significantly reduce complexity due to the lower number of channels and spatial resolution of network inputs.

\subsection{Wavelet Transform Block}
The DWT offers a powerful and compact way to represent image data by decomposing input image  $I\in \mathbb{R}^{C \times H \times W}$ into multiple frequency sub-bands at different resolutions, which can be denoted as
\begin{equation}
    WT(I)=\{LL^{({1})},LH^{({1})},HL^{({1})},HH^{({1})}\}.
\end{equation}
Multiple levels of decomposition can be employed by recursively applying a wavelet transform to the $LL$ sub-bands. In our framework, we employ 2-level 2D Haar DWT to decompose the input image, thus obtaining the following subbands: $\{LL^{({2})},LH^{({2})},HL^{({2})},HH^{({2})},LH^{({1})},HL^{({1})},HH^{({1})}\}$. The second-level sub-bands have a quarter of the original image resolution, i.e. $LL^{({2})},LH^{({2})},HL^{({2})} \in \mathbb{R}^{H/4 \times W/4 \times 3}$.
This decomposition enables the network to isolate and process coarse structures and fine details separately, which is particularly advantageous for deblurring tasks where sharp edges and texture preservation are critical. Moreover, it enables significant gains in terms of computational complexity for several reasons. First, the level-1 high frequency sub-bands are only marginally useful for the deblurring process as they contain limited residual information, so we can avoid processing them to save computations and skip them directly to the output. Moreover, in essence, the wavelet decomposition allows us to trade space and channels for each other, offering flexible tradeoffs for complexity. Since the level-2 sub-bands derived from $LL^{({1})}$ have been decimated by a factor of 4 in each dimension, all the operations whose complexity depends on spatial size, immediately require fewer operations. By processing $LL^{({2})},LH^{({2})},HL^{({2})},HH^{({2})}$ and concatenating their features along the channel dimension, the design allows flexible allocation of the dimensionality of feature space to save complexity. Indeed, we remark that co-located wavelet coefficients in different sub-bands are strongly correlated and learned projections along this dimension can easily exploit their joint information with a compact number of features.

More in detail, the subbands are processed independently by a convolutional layer and the results concatenated along the channel dimension: 
\begin{align}
    \mathbf{h}_{i} &= \text{Conv}_{i}(\mathbf{s}_i) \quad \mathbf{s}_i\in \{LL^{({2})},LH^{({2})},HL^{({2})},HH^{({2})}\} \\
    \mathbf{h} &= [\mathbf{h}_{1},\mathbf{h}_{2},\mathbf{h}_{3},\mathbf{h}_{4}] \in \mathbb{R}^{H/4 \times W/4 \times C}
\end{align}

In the decoder part, the output features are used to predict the level-2 wavelet coefficients, which are then combined with the original first-level high-frequency components and reconstructed into the spatial domain using the inverse wavelet transform.

\subsection{Efficient U-Net}

The main backbone of the model follows an encoder-decoder structure typical of several state-of-the-art image restoration models. Its input is the tensor $\mathbf{h}$ of processed level-2 wavelet sub-bands. 

The encoder comprises three hierarchical feature levels with increasing channels of $d=16,32,64$ and decreasing spatial resolution. Each level integrates one or more residual blocks that refine the feature representation. Each residual block processes an input feature map $\mathbf{z}$ according to:
\begin{equation}
    \mathbf{z'} = \mathbf{z} + C(\mathbf{z})
\end{equation}
where $C(\cdot)$ comprises two convolution layers, each with a kernel size of $3 \times 3$, 
the first layer activated using SiLU activation. Spatial downsampling between levels is performed using strided convolutions.

In the decoder, each level includes residual refinement blocks and upsampling layers, with skip connections concatenating corresponding encoder features. The depth information is injected via depth adapter modules, which transform the input depth map into a latent feature space and align it spatially with the encoder deepest representation. This depth feature modulates the decoder image features through a series of learned depth-guided prompting modules.
\begin{equation}
    \mathbf{z'}=A(\mathbf{z},\mathbf{d})
\end{equation}
where $\mathbf{d}$ denotes the depth map and $A(\cdot,\cdot)$ denotes the depth-adaptive fusion applied to decoder feature $\mathbf{z}$. To effectively integrate depth information into the visual feature stream, we design a lightweight adapter module that adaptively modulates image features using learned depth cues. In this setting, depth features serve as structural guidance, emphasizing edges and global layouts that aid in restoring sharpness and boundary. As illustrated in Fig. \ref{fig:adapter}, and inspired by the guided filtering framework \cite{he2012guided}, we adopt the approach of \cite{yi2024deep} by mimicking the second-order statistical behavior of guided filters. Specifically, the depth features are passed through two parallel convolutional branches that approximate the second-order statistics of the original guided filter, then passed through a sigmoid to stabilize the operations and then multiplied by the image features. The resulting depth-conditioned image features are concatenated with the original image features and processed by a lightweight convolutional layer followed by a simple channel attention mechanism from \cite{chen2022simple}, which enhances important feature channels while maintaining computational efficiency. Depth features processed by the attention operation are also propagated forward to the next stage, and possibly upsampled.

\section{Experiments}

\subsection{Experiments setting}
In our experiments, we follow the same setting as \cite{yi2024deep}. To be more specific, the dataset we use is a subset of the ARKitScenes dataset \cite{dehghan2021arkitscenes}, specifically the portion used for RGB-D guided upsampling, which contains 29,264 image-depth pairs in the training set. The validation set is composed by 500 pairs sampled randomly from the original validation set. Image blur is simulated by randomly choosing a blur kernel from a set of standard benchmark kernels, following the approach from \cite{5206815}. We remark that this choice of dataset is motivated by the availability of images with associated real depth maps produced by mobile Lidars.
The L1 loss and cosine similarity loss are used for training, which is run for about 400 epochs. Training has been performed with patches of size $256\times 256$ pixels. The initial learning rate is $10^{-4}$, and the Adam optimizer with $\beta_1=0.9$ and $\beta_2=0.999$ and cosine annealing decay policy are utilized.

\subsection{Main Result}

\begin{figure}
    \centering
    \includegraphics[width=\linewidth]{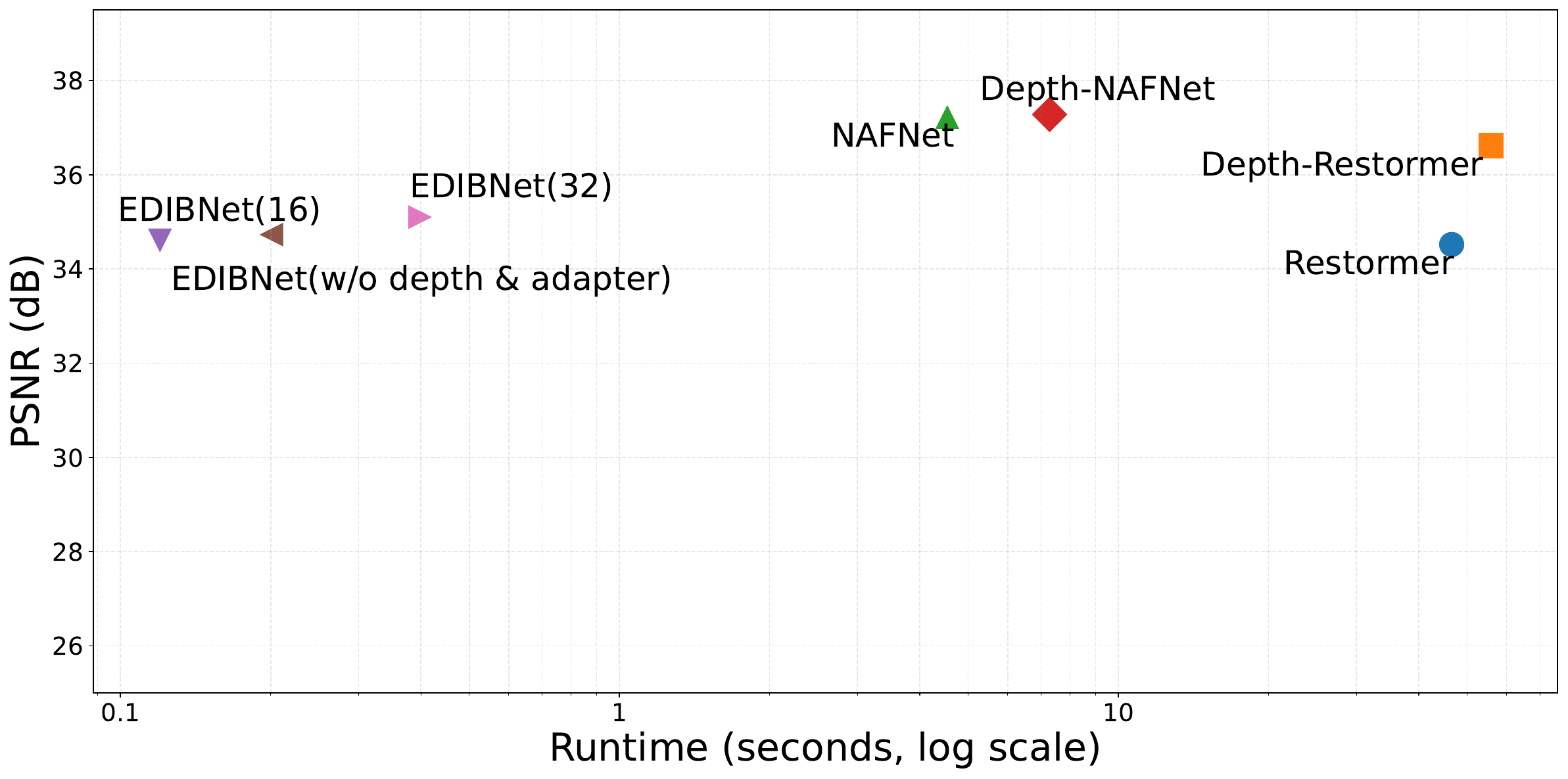}
    \caption{Deblurring image quality (PSNR) against Runtime (sec.).}
    \label{fig:performance_vs_runtime}
\end{figure}

\begin{figure*}[t]
  \captionsetup{font=small}
  \captionsetup[sub]{font=scriptsize,skip=2pt}
  \centering
  \begin{subfigure}[t]{0.23\linewidth}
    \includegraphics[width=\linewidth]{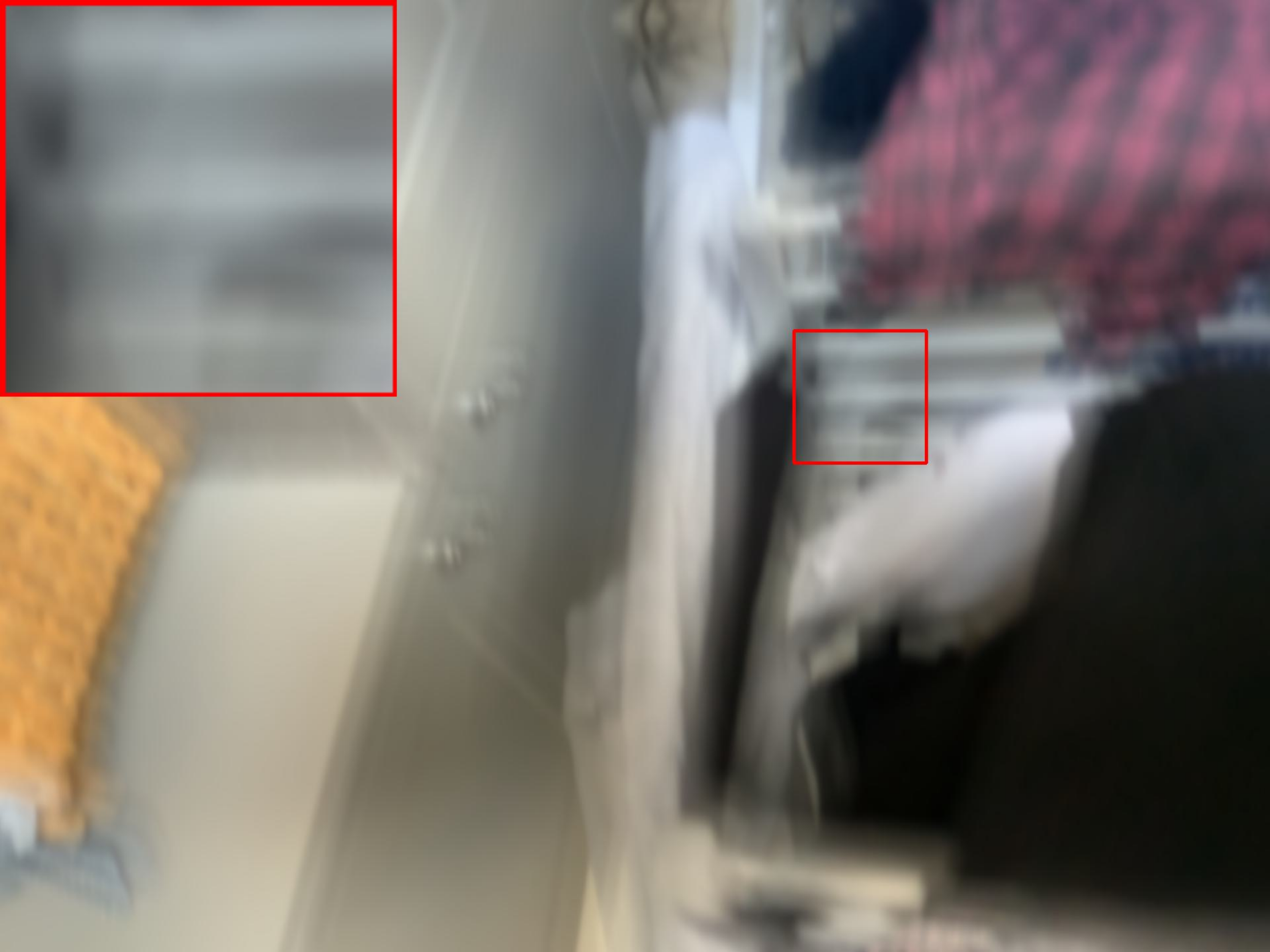}
    \caption{blur}
    \label{fig:45662975_26088_blur}
  \end{subfigure}
  \begin{subfigure}[t]{0.23\linewidth}
    \includegraphics[width=\linewidth]{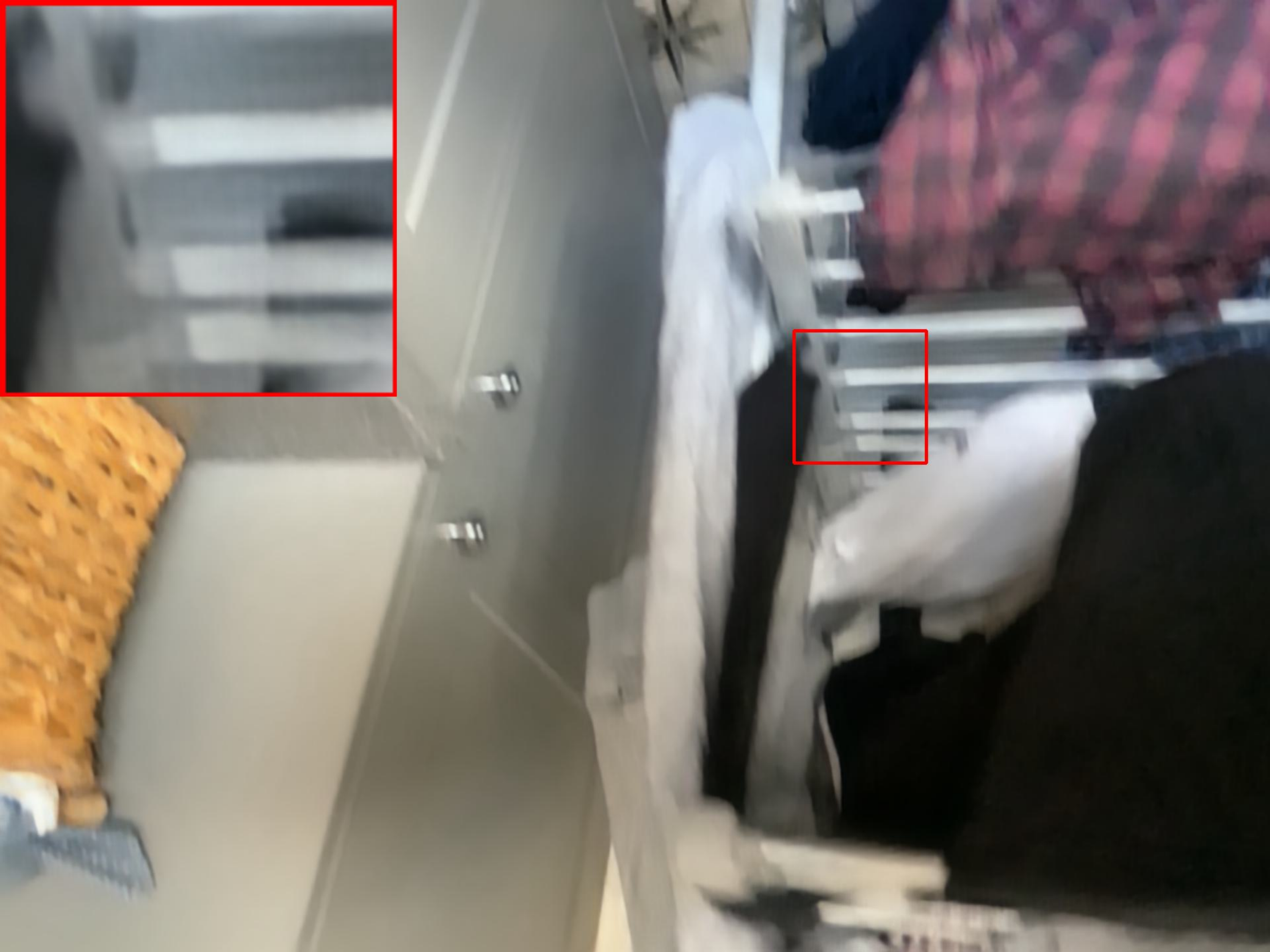}
    \caption{Restormer}
    \label{fig:45662975_26088_restormer}
  \end{subfigure}
  \begin{subfigure}[t]{0.23\linewidth}
    \includegraphics[width=\linewidth]{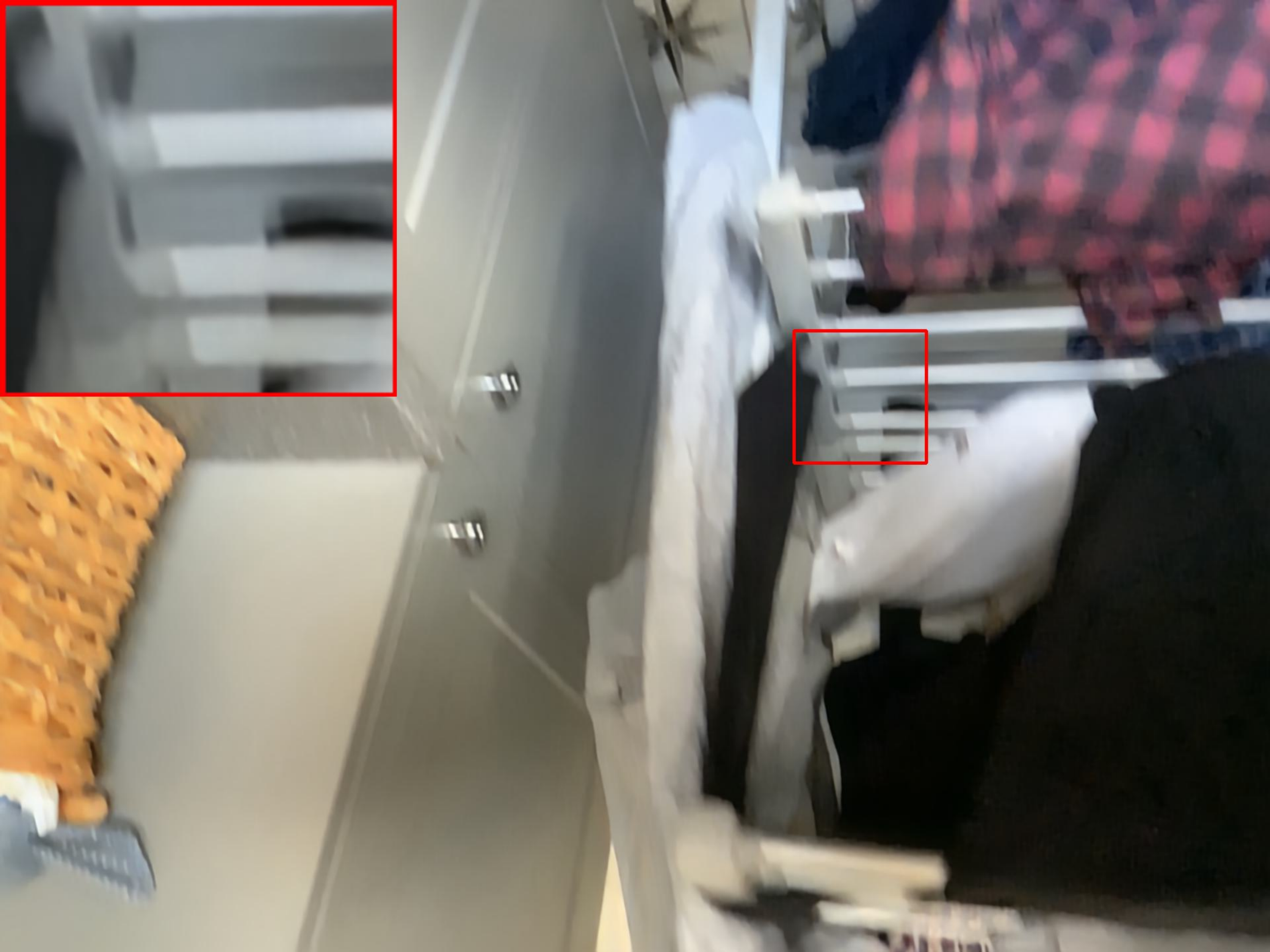}
    \caption{Depth Restormer}
    \label{fig:45662975_26088_depth_Restormer}
  \end{subfigure}
  \begin{subfigure}[t]{0.23\linewidth}
    \includegraphics[width=\linewidth]{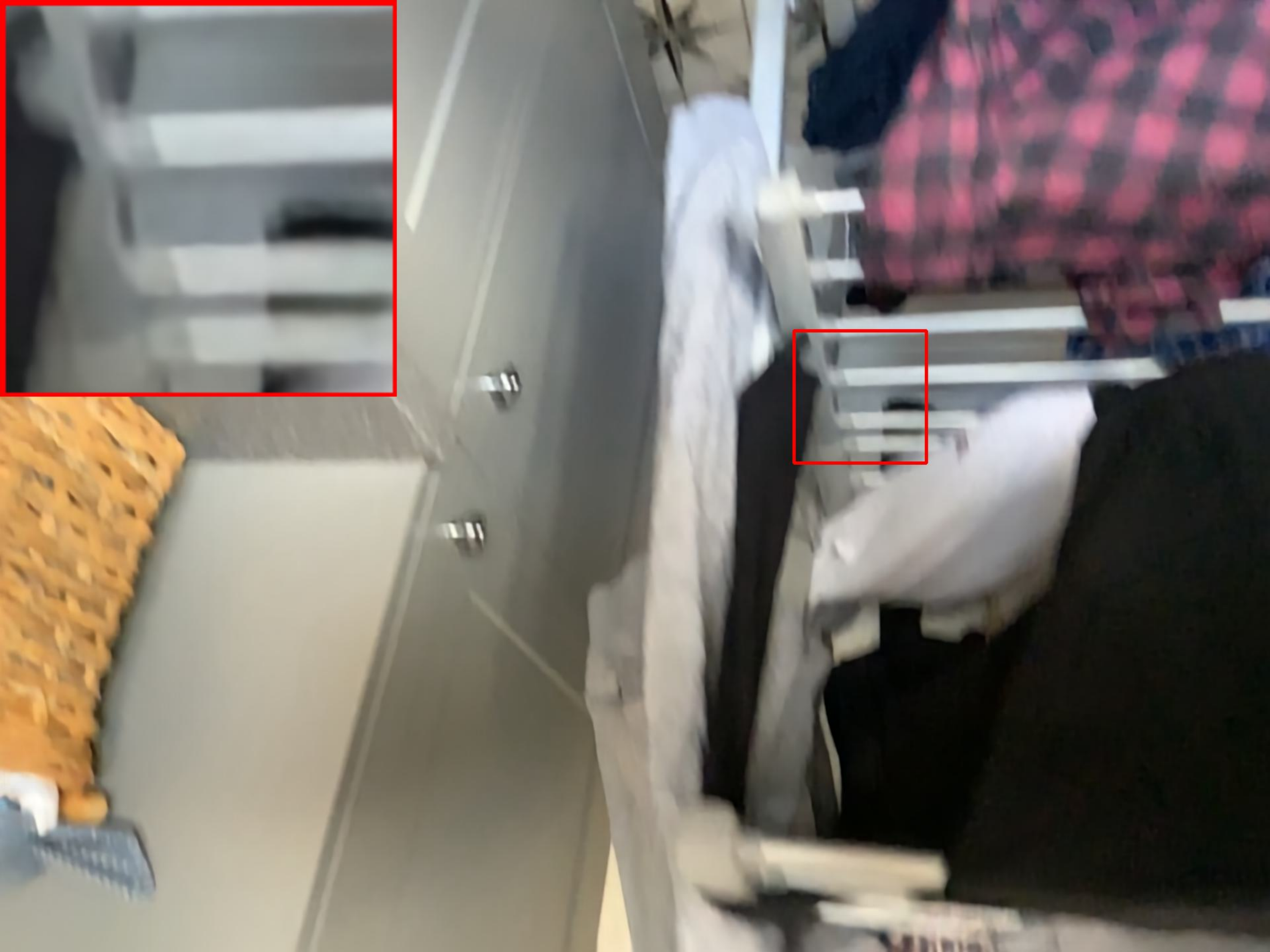}
    \caption{NAFNet}
    \label{fig:45662975_26088_NAFNet}
  \end{subfigure}
  
  \vspace{4pt}
  
  \begin{subfigure}[t]{0.23\linewidth}
    \includegraphics[width=\linewidth]{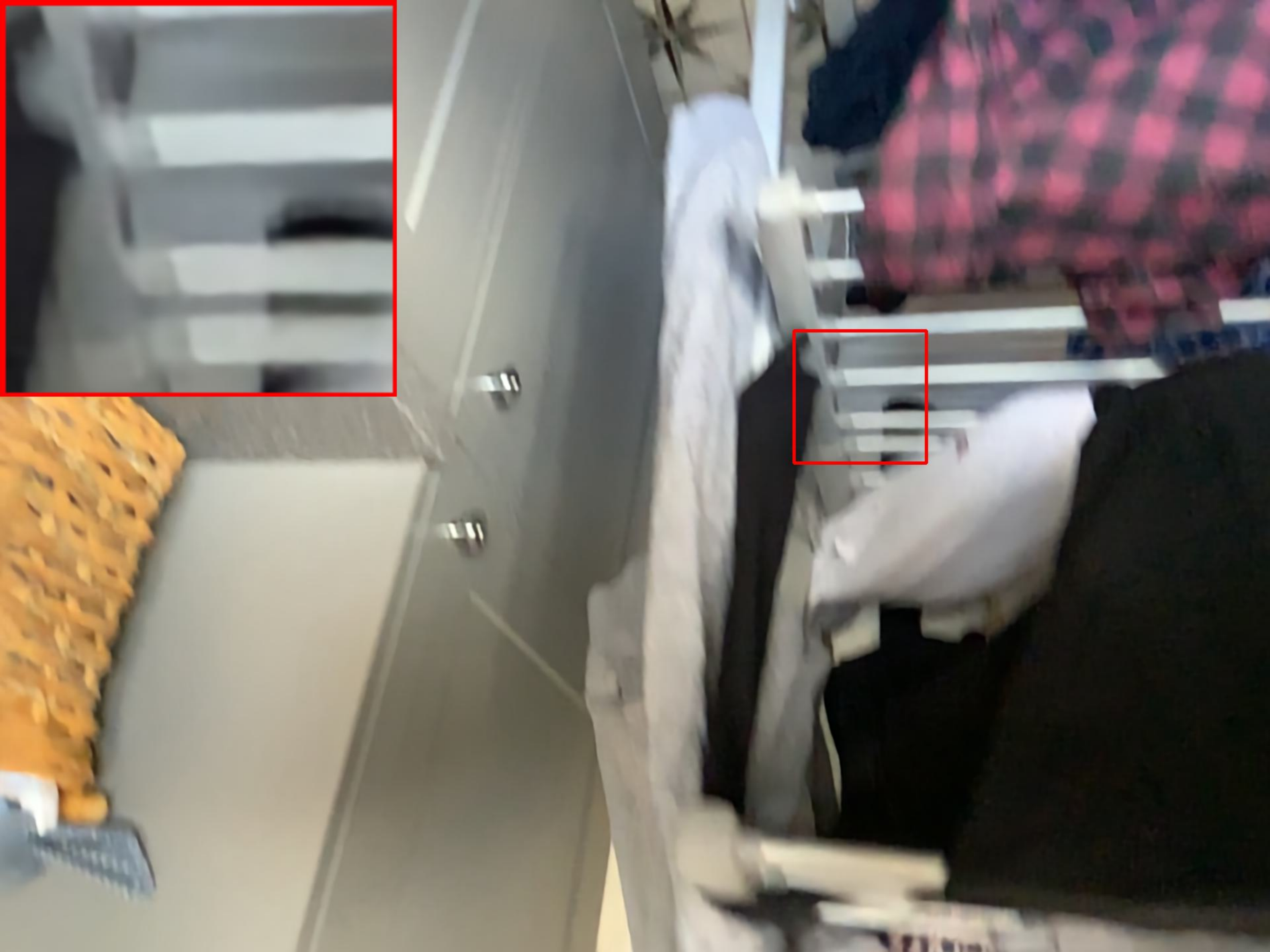}
    \caption{Depth NAFNet}
    \label{fig:45662975_26088_DepthNAFNet}
  \end{subfigure}
  \begin{subfigure}[t]{0.23\linewidth}
    \includegraphics[width=\linewidth]{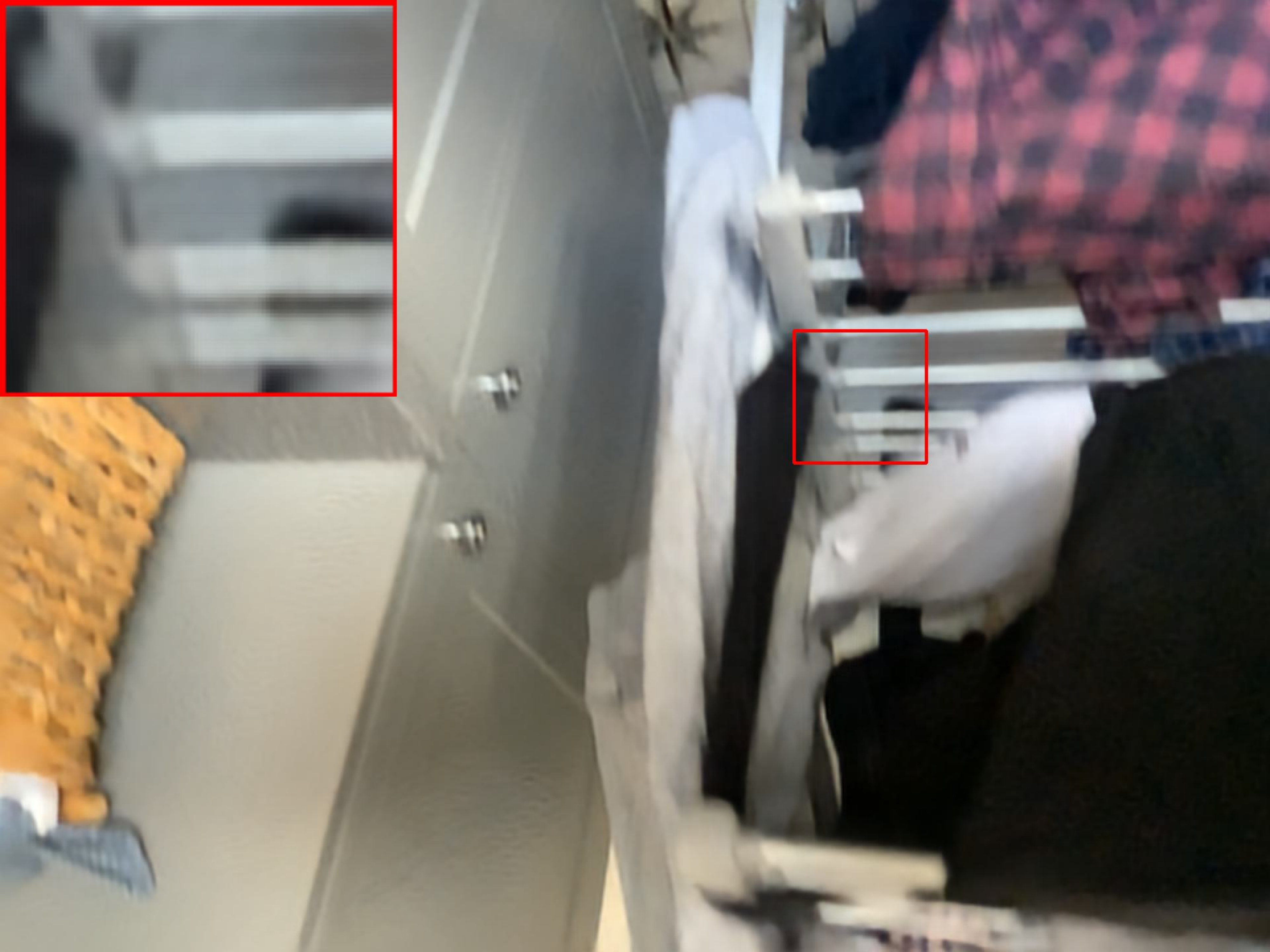}
    \caption{EDIBNet(w/o depth \& adapter)}
    \label{fig:45662975_26088_Wavelet}
  \end{subfigure}
  \begin{subfigure}[t]{0.23\linewidth}
    \includegraphics[width=\linewidth]{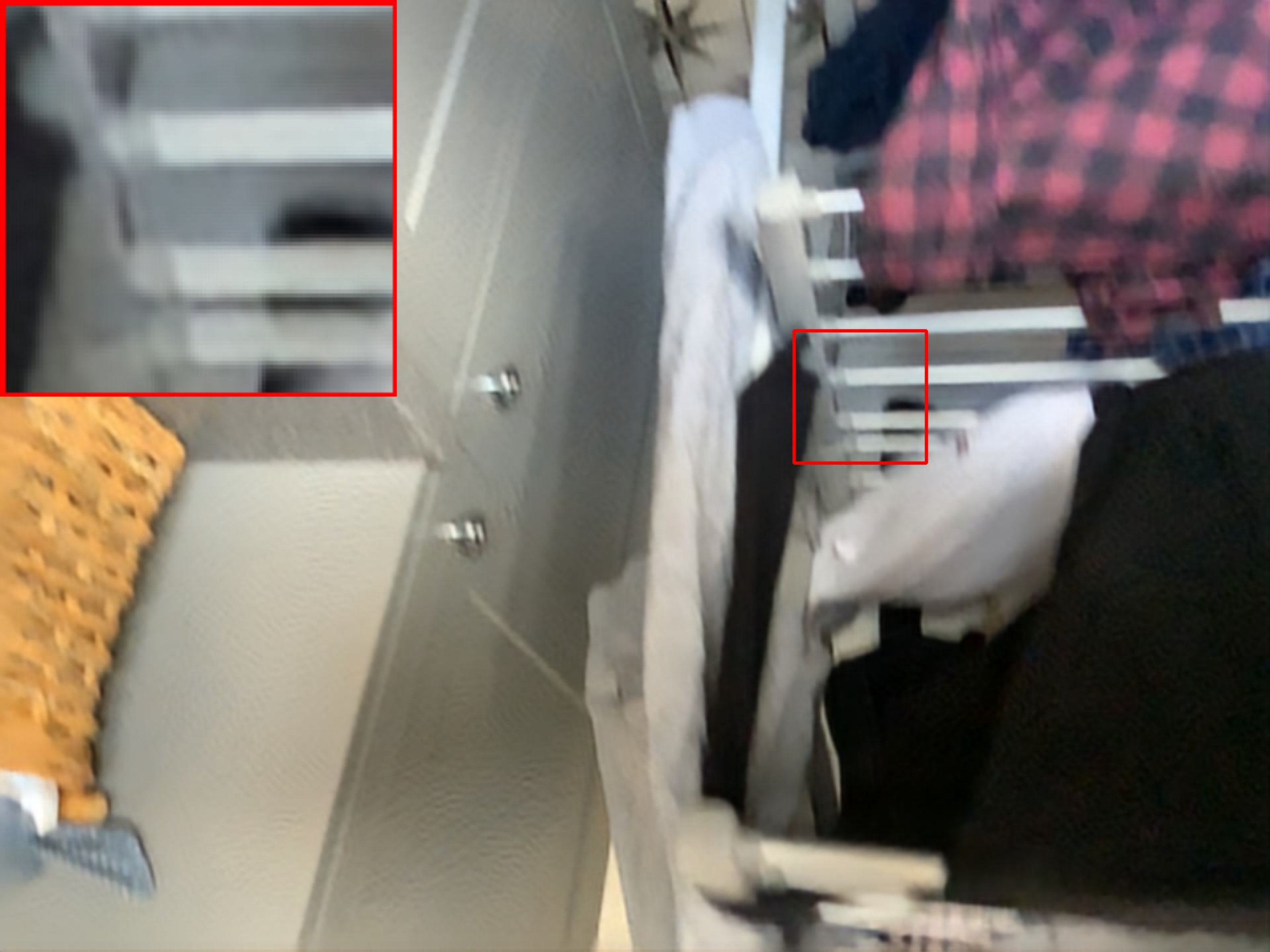}
    \caption{EDIBNet (channel=16) }
    \label{fig:45662975_26088_DepthWavelet}
  \end{subfigure}
  \begin{subfigure}[t]{0.23\linewidth}
    \includegraphics[width=\linewidth]{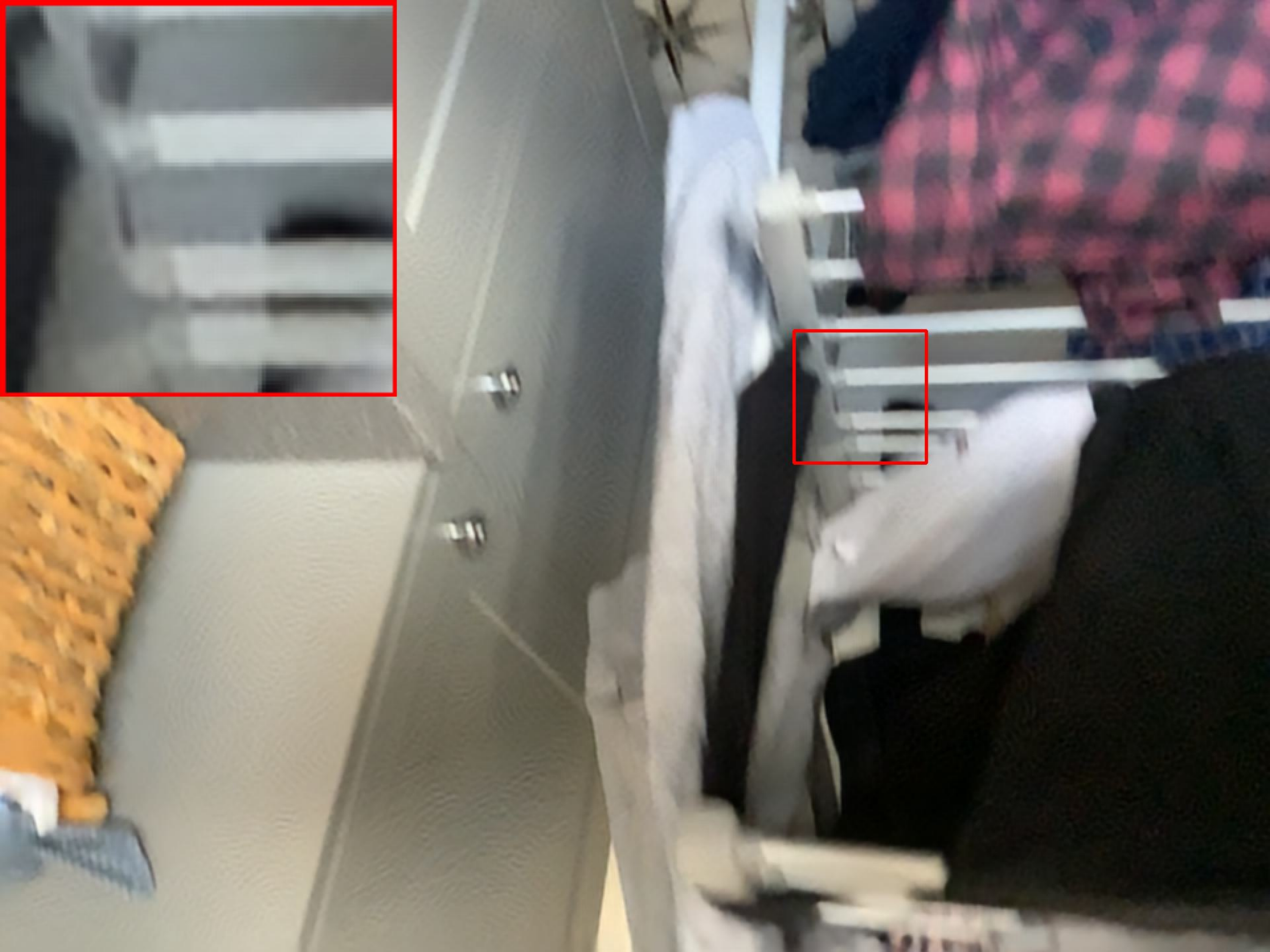}
    \caption{EDIBNet (channel=32) }
    \label{fig:45662975_26088_DepthWaveletMid}
  \end{subfigure}

  \begin{subfigure}[t]{0.23\linewidth}
    \includegraphics[width=\linewidth]{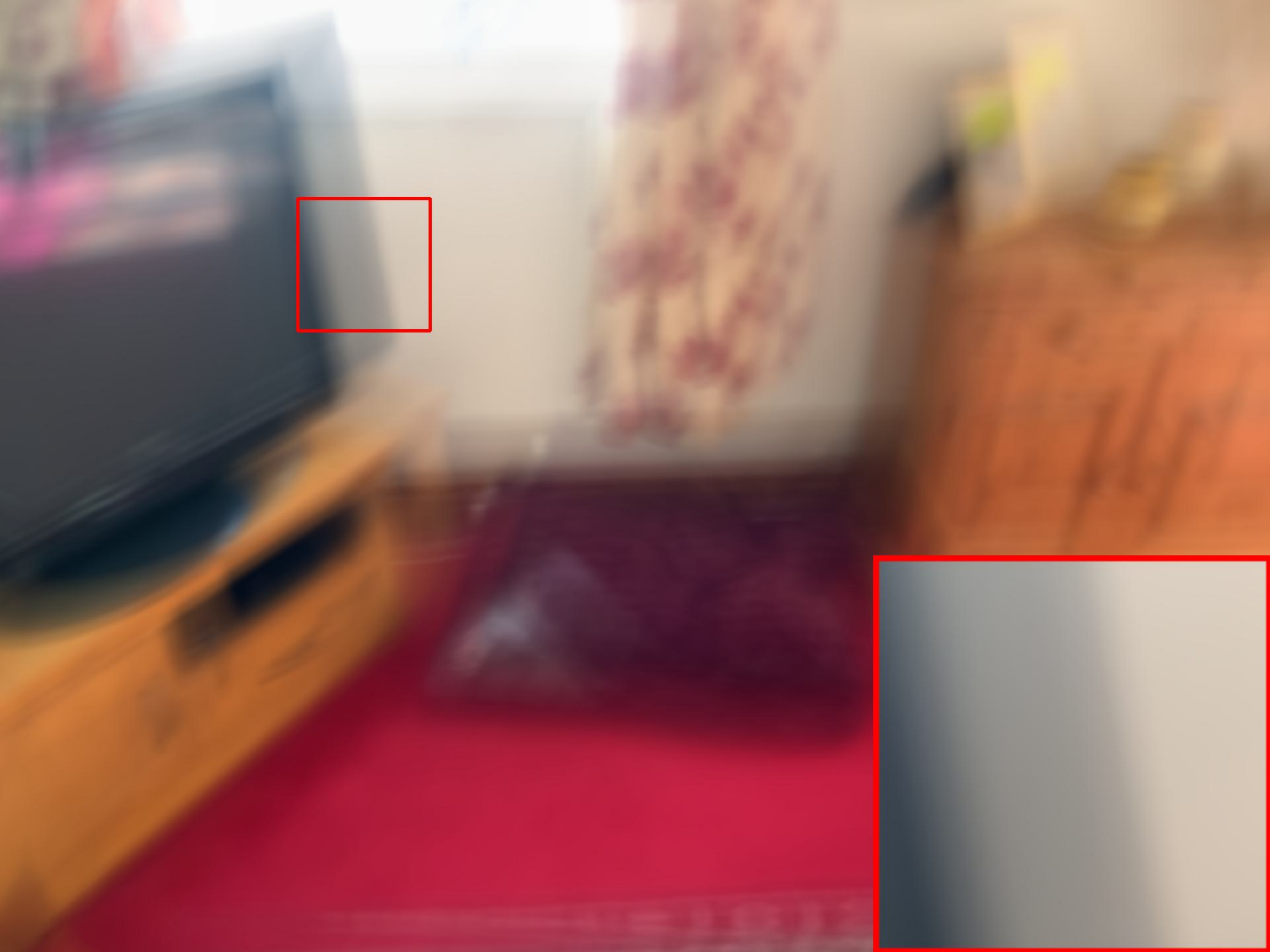}
    \caption{blur}
    \label{fig:45663149_63317.287_blur}
  \end{subfigure}
  \begin{subfigure}[t]{0.23\linewidth}
    \includegraphics[width=\linewidth]{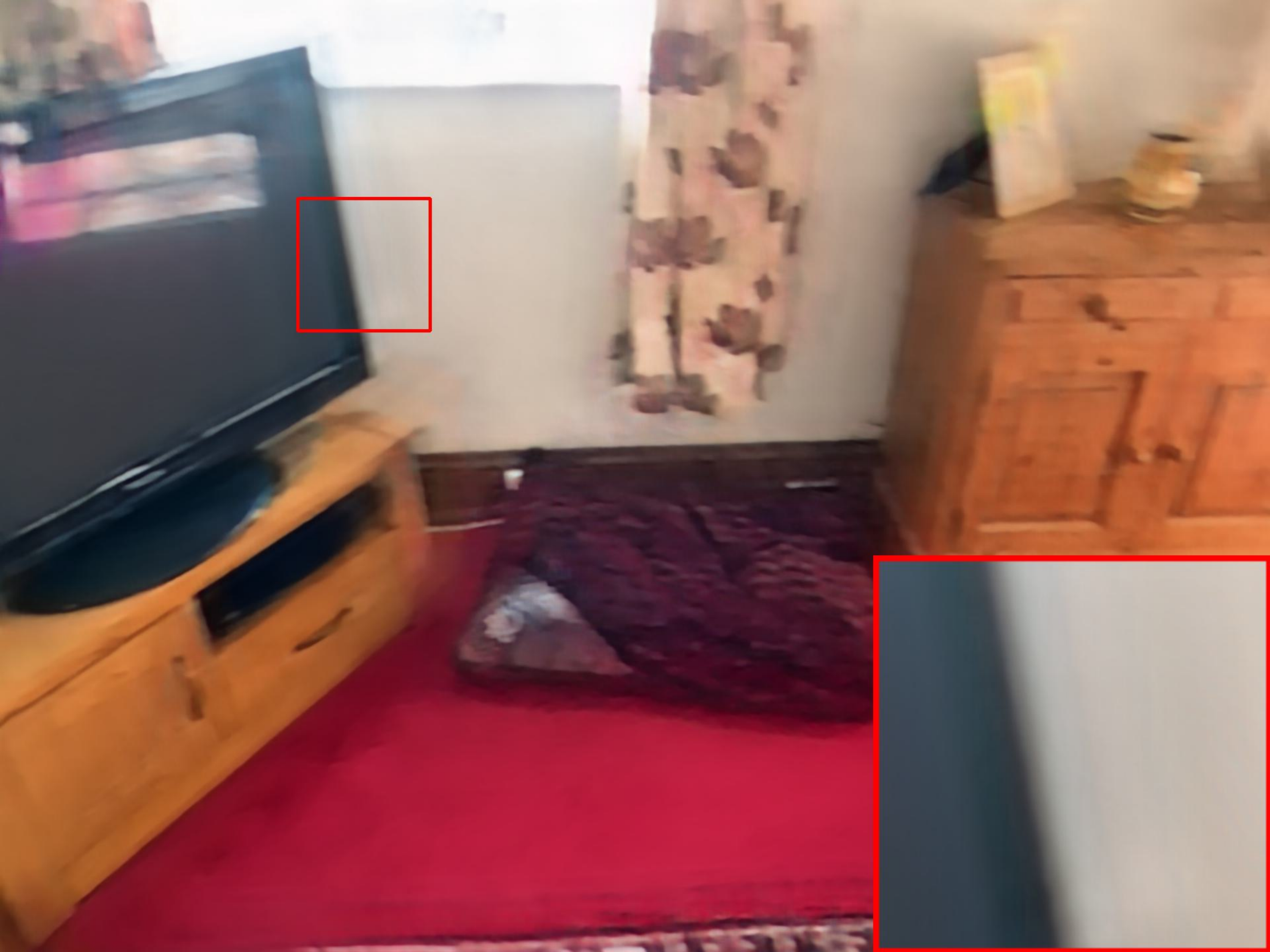}
    \caption{Restormer}
    \label{fig:45663149_63317.287_restormer}
  \end{subfigure}
  \begin{subfigure}[t]{0.23\linewidth}
    \includegraphics[width=\linewidth]{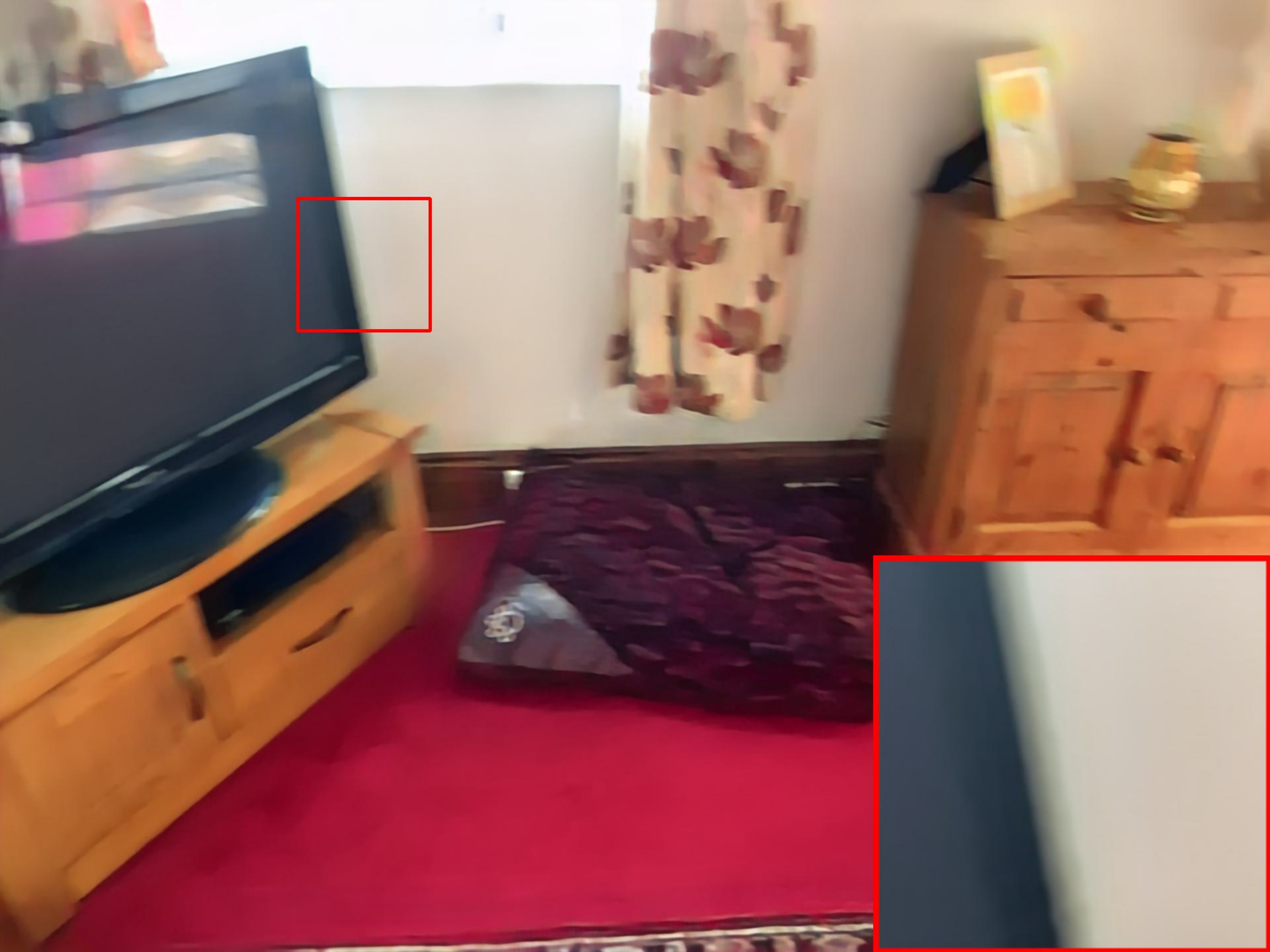}
    \caption{Depth Restormer}
    \label{fig:45663149_63317.287_depth_Restormer}
  \end{subfigure}
  \begin{subfigure}[t]{0.23\linewidth}
    \includegraphics[width=\linewidth]{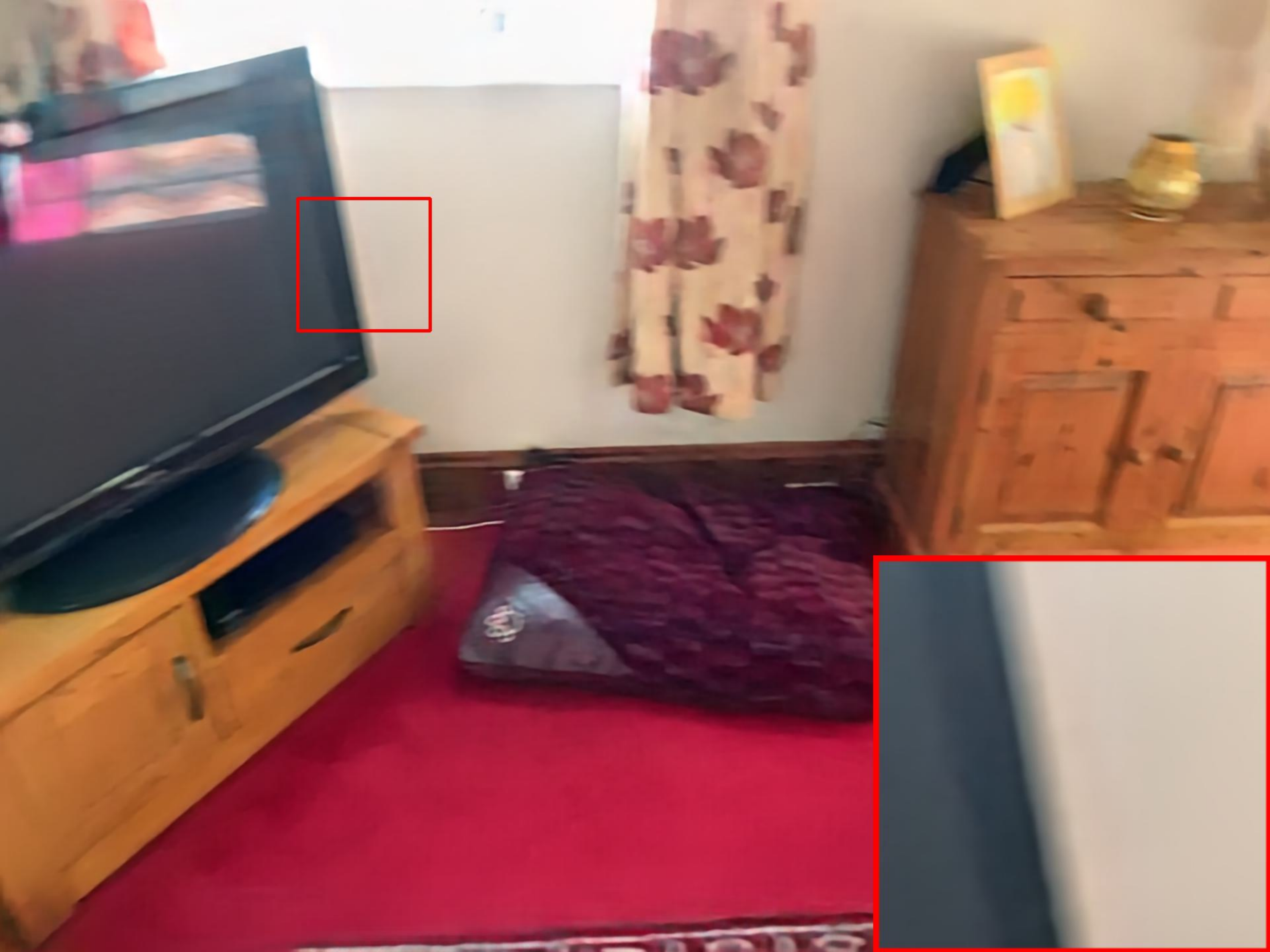}
    \caption{NAFNet}
    \label{fig:45663149_63317.287_NAFNet}
  \end{subfigure}
  
  \vspace{4pt}
  
  \begin{subfigure}[t]{0.23\linewidth}
  \includegraphics[width=\linewidth]{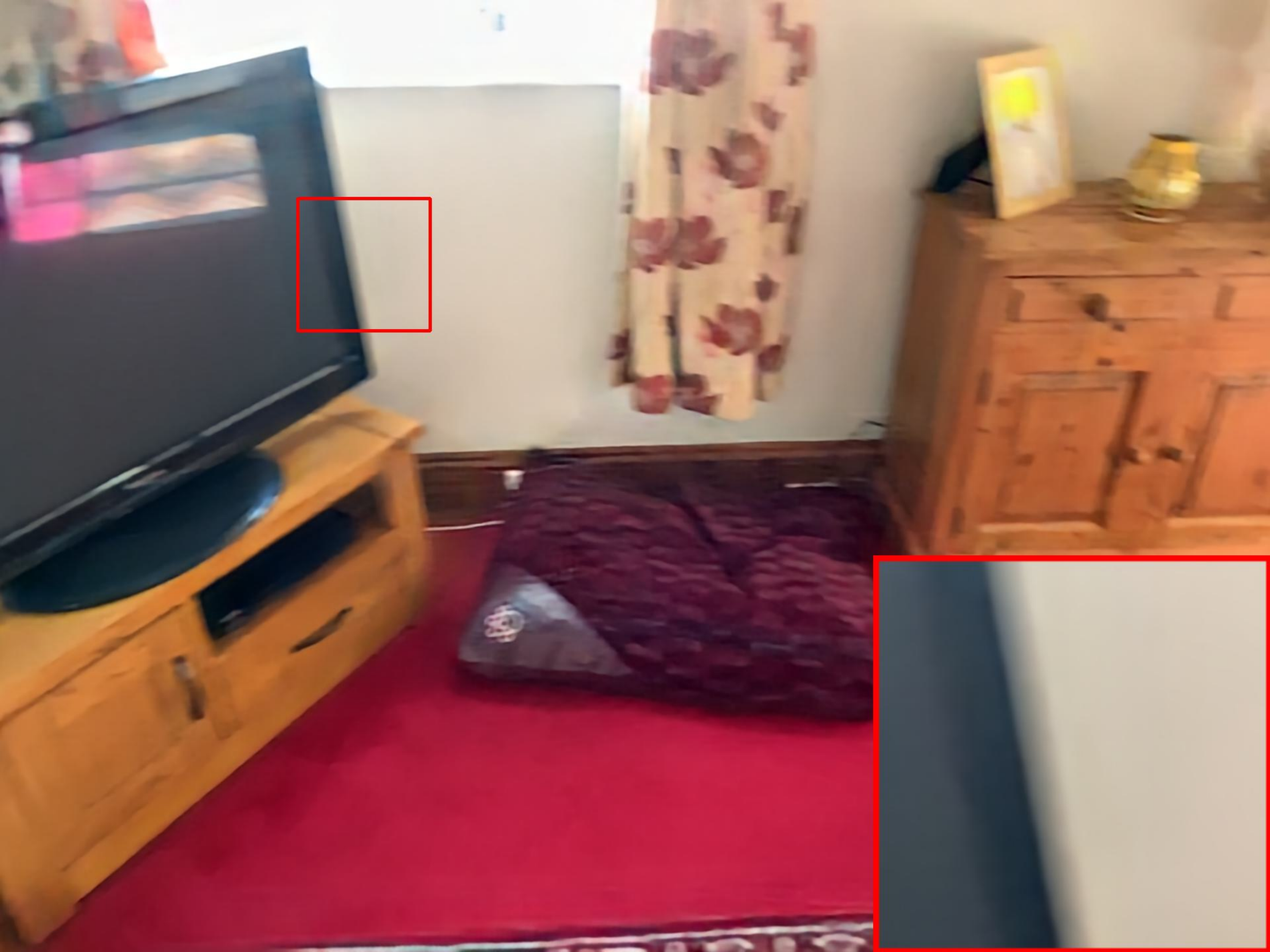}
    \caption{Depth NAFNet}
    \label{fig:45663149_63317.287_DepthNAFNet}
  \end{subfigure}
  \begin{subfigure}[t]{0.23\linewidth}
    \includegraphics[width=\linewidth]{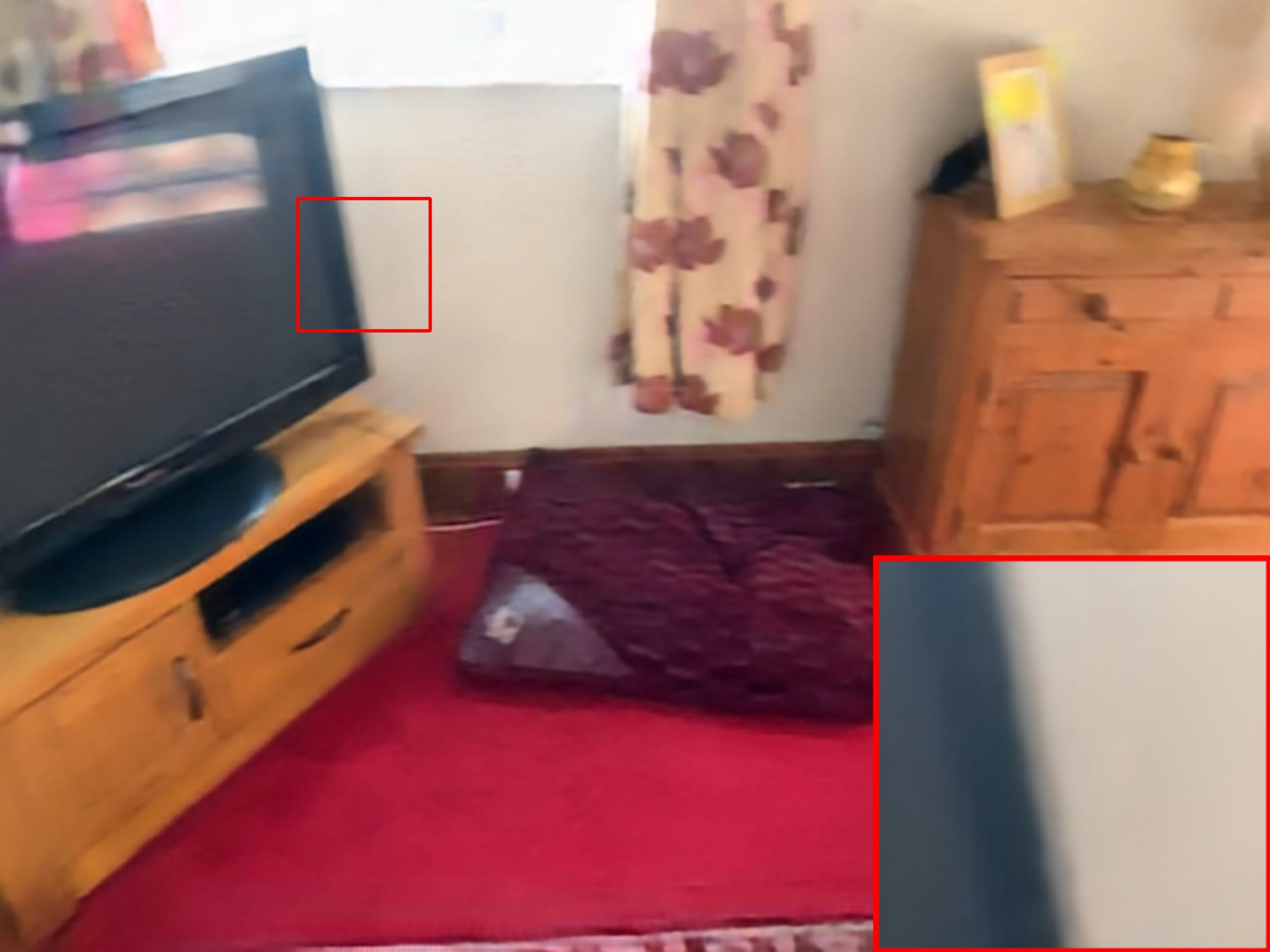}
    \caption{EDIBNet(w/o depth \& adapter) }
    \label{fig:45663149_63317.287_Wavelet}
  \end{subfigure}
  \begin{subfigure}[t]{0.23\linewidth}
    \includegraphics[width=\linewidth]{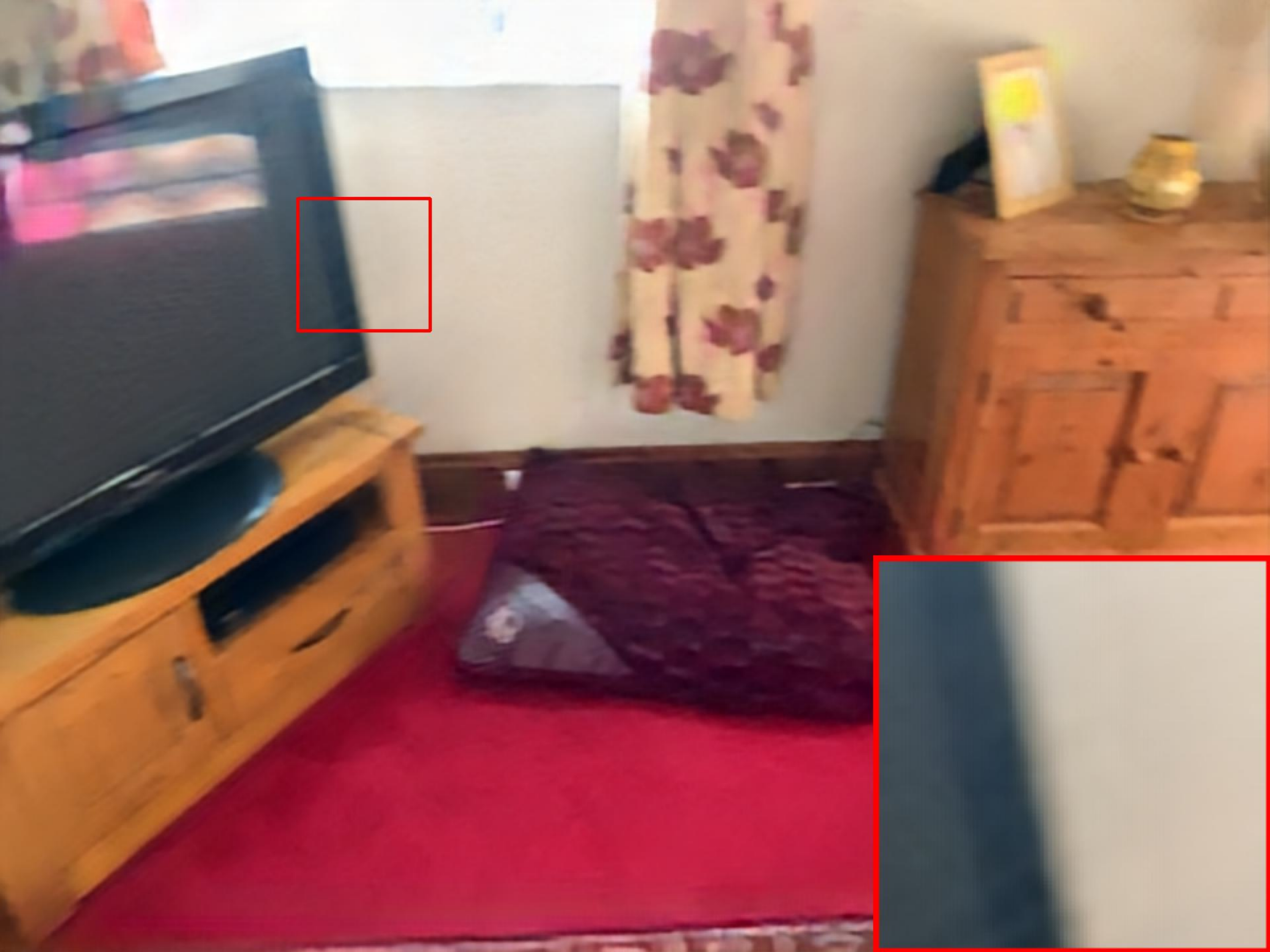}
    \caption{EDIBNet (channel=16)}
    \label{fig:45663149_63317.287_DepthWavelet}
  \end{subfigure}
  \begin{subfigure}[t]{0.23\linewidth} 
    \includegraphics[width=\linewidth]{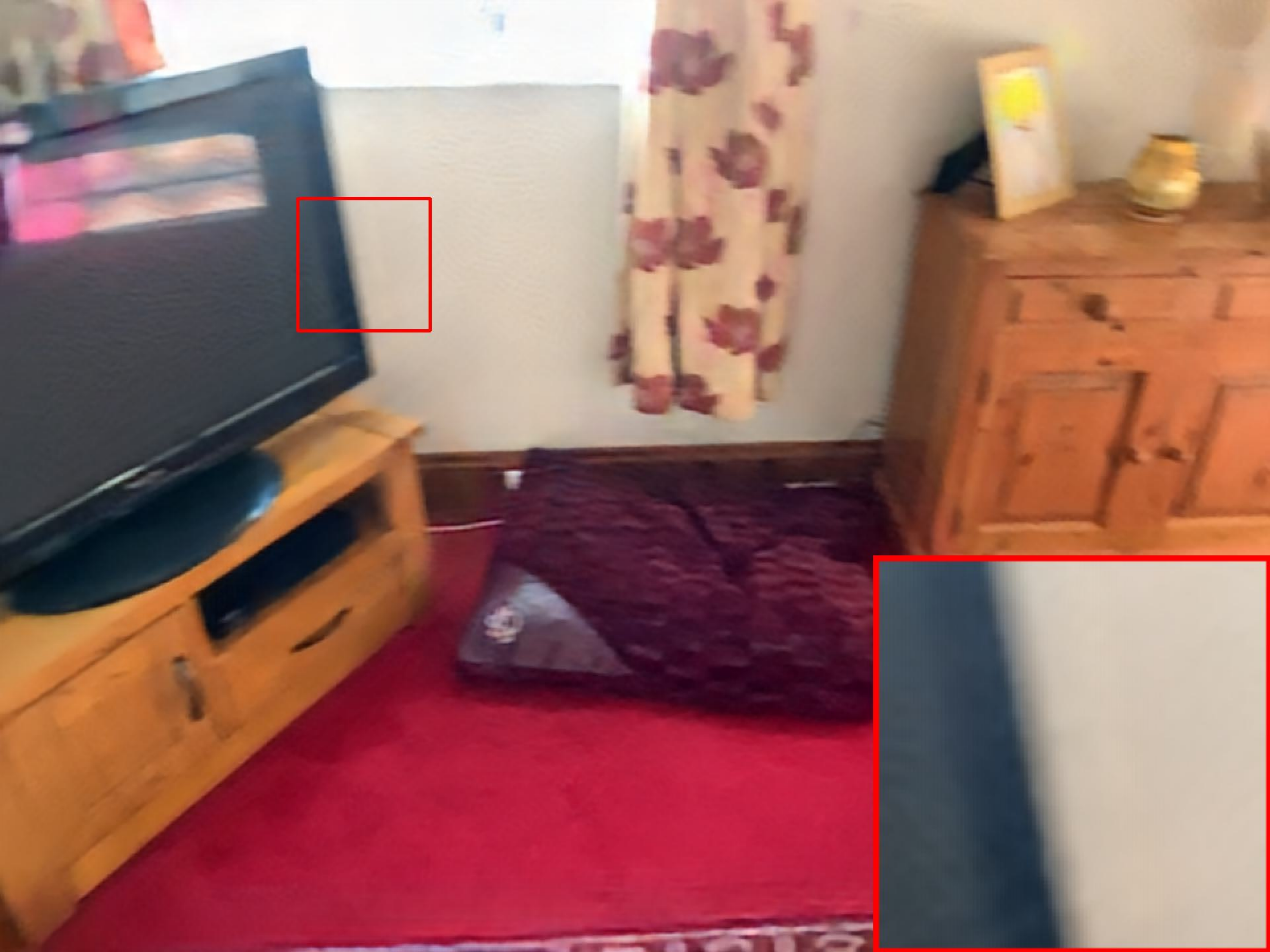}
    \caption{EDIBNet (channel=32)}
    \label{fig:45663149_63317.287_DepthWaveletMid}
  \end{subfigure}

\caption{Qualitative results.}
\label{fig:allresults}
  
\end{figure*}

\begin{table*}[t]

  \caption{Main deblurring results. Runtime is measured on an NVIDIA Jetson Orin Nano edge device, and FLOPs are computed using ARKitScenes resolutions: $1440 \times 1920$ for RGB images and $192 \times 256$ for depth maps.}

\centering
    \begin{tabular}{l ccccccc}
                \textbf{Model}                &\textbf{PSNR (dB) $\uparrow$}  &\textbf{SSIM$\uparrow$}  & \textbf{LPIPS$\downarrow$} & \textbf{Parameters (M)}&\textbf{FLOPs (G)}&\textbf{Runtime (s)}&\textbf{Memory (MB)}\\
    \hline\hline
                Restormer \cite{zamir2022restormer}   & 34.52                &0.9318        &0.2596 &26.1&4083&46.56&32456   \\
    \hline
                Depth-Restormer \cite{yi2024deep}    & 36.62                   & 0.9446   & 0.2223  &30.0&8786&55.84&41304\\
    \hline            
                    NAFNet \cite{chen2022simple}           & 37.24               &0.9430      &0.2474  & 17.1&673&4.54&4216\\
    \hline
                Depth-NAFNet \cite{yi2024deep}       & 37.28   & 0.9434   & 0.2433 &23.7 &1388&7.28 &11260
\\

    \hline
    \textbf{EDIBNet} (w/o depth \& adapter)     &34.59        &0.9667 & 0.3093 &1.45&15&0.12&280\\
    \hline
    \textbf{EDIBNet} (channel=16)               &34.73        &0.9673 &0.3117 &2.84 &44 &    0.20&358          \\
    \hline
    \textbf{EDIBNet} (channel=32) &35.10 &0.9681 &0.2971 &11.3 &178 &0.40&816\\
    \hline
    \end{tabular}
\label{table:main}
\end{table*}

The main results of our experiments are presented in Table~\ref{table:main}, which compares our method against SOTA approaches trained on the same dataset. It is evident that our network achieves competitive performance relative to its compact size. Notably, all the models outperform Restormer in terms of PSNR. When real-world depth information is integrated, our model gains an additional 0.14 dB in PSNR, with a marginal runtime increase of less than 0.08 s. We remark how there are approximately two orders of magnitude between Restormer and the proposed EDIBNet in terms of FLOPs, runtime and memory. This confirms the efficiency and high parallelism of our depth-adaptive design. We remark that for those models exceeding the 8GB memory limit of the Jetson Orin Nano, runtime accounts for image tiling and sequential evaluation, highlighting the difficulty of running expensive SOTA models on edge platforms.

Meanwhile, Fig.\ref{fig:performance_vs_runtime} provides a visual depth of the runtime-quality tradeoff offered by various methods. We can easily notice that all the variants of the proposed EDIBNet method are in the top-left corner, indicating significantly better latency and good image quality. 

Qualitative comparisons shown in Fig.~\ref{fig:allresults} further validate the effectiveness of our method at recovering sharp edges and preserving fine details. Overall, these results demonstrate that our approach achieves an excellent balance between accuracy and efficiency, making it well-suited for image restoration on edge devices.

\subsection{Ablation Study}
In the ablation study, we carefully analyze our design decisions to validate their effectiveness. This study concerns the three main points of contribution of this paper, namely the wavelet decomposition level, wavelet types and the adapters architecture. All the ablation results use the variant with 16 feature channels as baseline.  

\subsubsection{Impact of Wavelet Decomposition Level}
In this section, we discuss the impact of the wavelet decomposition level, by comparing processing of $LL^{(1)}$ (1-level decomposition) instead of the proposed $LL^{({2})}$,$LH^{({2})}$,$HL^{({2})}$,$HH^{({2})}$ (2-level decomposition). We also compare again 3-level decomposition where $LL^{({2})}$ is further decomposed and all the high frequency subbands from the first and second level are skipped, as this would further reduce complexity.
As is shown in Table \ref{tab:level_comparison}, the proposed 2-level wavelet decomposition achieves the best overall performance, with a PSNR of 34.73 dB and SSIM of 0.9673, outperforming both the 1-level and 3-level variants. Besides the higher cost in FLOPs and Runtime, the image quality in 1st-level suffers from the insufficient frequency separation because it only performs a single round of decomposition, resulting in relatively coarse separation between low- and high-frequency components, which limits the network's ability to isolate blur artifacts from structural image content. On the other hand, the 3-level configuration yields the fastest runtime and the lowest FLOPs, but at the cost of noticeable performance degradation, likely due to excessive information loss from deeper decomposition. These results demonstrate that 2-level wavelet decomposition strikes the optimal balance between accuracy and efficiency.

\begin{table}[t]
\centering
\caption{Comparison between 1-level, 2-level and 3-level wavelet transform configurations and pixel.}
\setlength{\tabcolsep}{3.6pt}
\label{tab:level_comparison}
\begin{tabular}{lccc}
\toprule
\textbf{Metric} & \textbf{1-Level Wavelet} & \textbf{2-Level Wavelet} & \textbf{3-Level Wavelet}  \\
\hline
\midrule
PSNR (dB)       & 33.59                    &     \textbf{34.73} &32.47             \\  
\hline
SSIM            & 0.9639                    & \textbf{0.9673}  &0.9543                \\
\hline
Parameters (M)  & 2.84                     &2.84 &2.84  \\
\hline
FLOPs (G)       & 162.55                    &  44.81&15.59                      \\
\hline
Runtime (s)    &  0.68               &   0.20 &0.09                   \\
\hline
\bottomrule
\end{tabular}
\end{table}

\subsubsection{Impact of Wavelet basis}
In this section, we conclude our analysis on the wavelet transform by assessing the performance of additional wavelet bases. Table \ref{tab:wavelet_comparison}  provides a comparative evaluation of the Haar, \textbf{rbio1.1}, and \textbf{bior1.1} wavelets. Standard metrics including PSNR, SSIM, parameter count, FLOPs, and runtime are evaluated. We can notice that there is no significant difference between different wavelet bases. Notably, all configurations maintain the same number of parameters  FLOPs, and inference runtime, indicating that the choice of wavelet basis impacts performance quality without affecting computational efficiency. The Haar basis is therefore desirable for its ease of implementation.

\begin{table}[t]
\centering
\caption{Comparison between different types of wavelet transform basis.}
\label{tab:wavelet_comparison}
\begin{tabular}{lccc}
\toprule
\textbf{Metric} & \textbf{Haar}& \textbf{rbio1.1}& \textbf{bior1.1}\\
\hline
\midrule
PSNR (dB)       & 34.73     & 34.49  &   34.42            \\
\hline
SSIM            & 0.9673    & 0.9649  &    0.9658            \\
\hline
Parameters (M)  & 2.84      &2.84&  2.84                 \\
\hline
FLOPs (G)       & 44.81     &44.81&    44.81              \\
\hline
Runtime (ms)    & 0.20     & 0.20    &  0.20       \\
\hline
\bottomrule
\end{tabular}
\end{table}

\subsubsection{Impact on adapters}

\begin{table}[t]
\centering
\caption{Comparison between adapters.}
\label{tab:adapter_comparsion}
\begin{tabular}{lcc}
\toprule
\textbf{Metric} & \textbf{Our adapter} & \textbf{adapter from\cite{yi2024deep}}   \\
\hline
\midrule
PSNR (dB)       & 34.73                    & \textbf{34.82}                \\
\hline
SSIM            & \textbf{0.9673}                    & 0.9662                  \\
\hline
Parameters (M)  & \textbf{2.84}                     & 3.07                      \\
\hline
FLOPs (G)       & \textbf{44.81}                    & 51.34                   \\
\hline
Runtime (s)    & \textbf{0.20}                    &0.30                   \\
\hline
\end{tabular}
\end{table}

In this section, we evaluate the effectiveness of the design of the depth adapter. We already presented in Table \ref{table:main} that the adapter and depth information improve image quality. Here we focus on a comparison with the adapter presented in \cite{yi2024deep} which was already quite effective and lightweight but has been further streamlined in this work, as presented. Results are shown in Table~\ref{tab:adapter_comparsion}. While the PSNR of our adapter is marginally lower than that of \cite{yi2024deep}, our model achieves a higher SSIM, indicating better structural preservation. More importantly, the current design is more efficient: it reduces the number of parameters by over 7.5\% (2.84M vs. 3.07M), cuts FLOPs by approximately 12.7\% (44.81G vs. 51.34G), and lowers runtime from 0.20 ms to 0.15 ms. These improvements demonstrate that our adapter achieves comparable or better performance while being lighter, faster, and more suitable for deployment in resource-constrained environments.

\section{Conclusion}
We proposed an efficient neural network model for depth-guided image deblurring. By processing the blurred image in the wavelet domain, significant savings in terms of computational complexity can be achieved with a suitable lightweight encoder-decoder architecture while retaining image quality. Moreover, lightweight depth adapter modules effectively and efficiently combine depth information to improve image quality. The experiments demonstrate that our method achieves a favorable balance between performance and efficiency, outperforming larger models in runtime and computational resources by two orders of magnitude, while delivering competitive or superior image quality.

\bibliographystyle{IEEEtran}

\end{document}